\documentclass[final,1p,times]{elsarticle}
\usepackage{amssymb, amsmath}
\usepackage{graphicx,epsf}
\usepackage{color}

\journal{Nuclear Physics B}

\begin{document}

\begin{frontmatter}



\title{Sachdev-Ye-Kitaev Model as  Liouville Quantum Mechanics}

\author[label1]{Dmitry Bagrets}
\author[label1]{Alexander Altland}
\author[label2]{Alex Kamenev}
\address[label1]{Institut f\"ur Theoretische Physik, Universit\"at zu K\"oln,
Z\"ulpicher Stra\ss e 77, 50937 K\"oln, Germany}
\address[label2]{W. I. Fine Theoretical Physics Institute and School of Physics and Astronomy, University
of Minnesota, Minneapolis, MN 55455, USA}

%\author{}
%\address{}

\begin{abstract}
We show that the proper inclusion of soft reparameterization modes in the
Sachdev-Ye-Kitaev model of $N$ randomly interacting Majorana fermions reduces its
long-time behavior to that of Liouville quantum mechanics. As a result, all zero
temperature correlation functions decay with the universal  exponent $\propto
\tau^{-3/2}$ for times larger than the inverse single particle level spacing $\tau\gg
N\ln N$. In the particular case of the single particle Green function this behavior
is manifestation of the zero-bias anomaly, or scaling in energy as $\epsilon^{1/2}$.
We also present exact diagonalization study supporting our conclusions.
\end{abstract}

%\begin{keyword}
%Majorana fermions %\sep keyword



%\end{keyword}

\end{frontmatter}


\section{Introduction and the model}
\label{sec:Intro}

The Sachdev-Ye-Kitaev (SYK) model~\cite{Sachdev:2015, Kitaev:2015} is a system of $N$ Majorana fermions, $ \chi_j$, $j=1,\ldots,N$,  subject to a  four-fermion interaction,
\begin{equation}
\label{eq:model}
\hat H={1\over 4!}\sum\limits^N_{ijkl}J_{ijkl}\, \chi_i\chi_j\chi_k\chi_l\, , 
\end{equation}
where the coupling constants, $J_{ijkl}$, are independent Gaussian random variables distributed as
\begin{equation}
\label{eq:distribution}
\langle J_{ijkl}\rangle =0 \;; \quad\quad\quad \langle (J_{ijkl})^2\rangle =3!J^2/N^3.
\end{equation}
This seemingly innocent model has recently been recognized~\cite{Kitaev:2015,Polchinski:2016,Maldacena:2016} as a possible shadow of a two-dimensional gravitational bulk and the ensuing perspective to explore a holographic correspondence of lowest possible dimension in concrete terms.  The holographic principle as much as all other relevant features of the system are rooted in its exceptionally high level of symmetries. When formulated in the language of a (0+1)-dimensional `field theory' 
%(i.e. a Feynman path integral) 
the system shows approximate symmetry under reparameterization of time $\chi_j(\tau)\to \chi_j(f(\tau))$. In the limit of asymptotically slow time variations (compared to the characteristic interaction strengths $J$) that symmetry becomes exact. The infinite dimensional group of time reparameterizations is generated by a Virasoro algebra and for this reason the theory has been dubbed a `nearly conformally invariant' (NCFT) theory, and the emerging holographic principle an $NAdS_2/NCFT_1$ correspondence.

Thanks to the presence of the large parameter $N$ and the `infinite rangedness' of the interaction the path integral over Majorana configurations can be processed by stationary phase methods. It turns out that the mean field, physically equivalent to a self-consistent approximation for an interaction self-energy, spontaneously breaks the conformal symmetry down to the three-dimensional group $\mathrm{SL}(2,R)$ of conformal transformations in one dimension. This sets the stage for the emergence of an infinite dimensional manifold of Goldstone modes described by reparameterization of time, $\tau\to f(\tau)$, modulo ordinary conformal invariance. The fluctuations of these modes are damped  by  explicitly symmetry breaking time derivatives which enter the theory in combinations $\sim \partial_\tau/J$ and play a role comparable to that of an external magnetic field in a ferromagnet. Due to the low dimensionality of the model  all time-dependent correlation functions must be qualitatively affected by these fluctuations in the long time limit, where the breaking of symmetry becomes weak. In particular one should expect the mean-field amplitudes of all observables  (such as e.g. Green functions) to be qualitatively changed at low frequencies, i.e. in the limit where the Mermin-Wagner theorem enforces a restoration of the full symmetry via proliferating Goldstone mode fluctuations. 

Previous work \cite{Maldacena:2016} took a step towards  an effective  action $S[f(\tau)]$ for the
reparameterization Goldstone modes and studied their effects to lowest (quadratic)
order in perturbation theory. While this is sufficient to describe short time
correlations,  a different treatment is required in the infrared regime. In this paper we point out that the action $S[f]$ can be mapped
onto the action of the \emph{Liouville quantum mechanics}, i.e. quantum mechanics in an
exponential potential~\cite{Zamolodchikov:1996, Teschner:2001, Nakayama:2004}. This reformulation sets the stage for the
treatment of the correlation functions in fluctuation dominated long time regime. We will show that at long times,  $\tau J > N \ln N$, all correlation functions  crossover to 
qualitatively different power laws. The  ensuing long
time asymptotics imply the vanishing of the mean-field two-point function at low frequencies. Where the four point function is concerned they represent operator correlations beyond those perturbatively (diagrammatically) described previously. Finally, the equivalence of the low frequency problem to Liouville quantum mechanics implies~\cite{Shelton:1998, Balents:1997} that two-time correlation functions $\langle \mathcal{O}(\tau)\mathcal{O}(\tau')\rangle \sim |\tau-\tau'|^{-3/2}$ of arbitrary operators must decay with a universal $3/2$-power law at large times. It stands to reason that this type of universality must reflect in the manifestations of a holographic correspondence in the infrared. However, the discussion of such correspondences is beyond the scope of the present paper and may become a subject of future work.

\section{Preliminaries}

To proceed with the model defined by Eqs.~(\ref{eq:model}) and (\ref{eq:distribution}), we introduce 
replicas, labeled as $a=1,2,\ldots,n $, and average over $\{J_{ijkl}\}$ to find: 
\begin{equation}
\left\langle e^{-\sum\limits^n_{a=1} \int d\tau H^a}\right\rangle = e^{ \frac{J^2}{8N^3} \sum\limits^n_{a,b} \int d\tau  d\tau' \sum\limits^N_{ijkl}
\chi_i^a(\tau)\chi_j^a(\tau)\chi_k^a(\tau)\chi_l^a(\tau)\times \chi_i^b(\tau')\chi_j^b(\tau')\chi_k^b(\tau')\chi_l^b(\tau')}=
e^{ N \frac{J^2}{8} \sum\limits^n_{a,b} \int d\tau d\tau' \big[G^{ab}_{\tau,\tau'}\big]^4},
\end{equation}
where
\begin{equation}
G^{ab}_{\tau,\tau'}=- N^{-1}\sum\limits_i^N \chi_i^a(\tau)\chi_i^b(\tau').
\end{equation}
%Notice that $G^{aa}_{\tau,\tau}=1/2$. 
The field $G^{ab}_{\tau,\tau'}$ may be made dynamical by inserting unity in the partition function as
\begin{equation}
1=\int \mathcal{D}G\,\, \delta\Big( N G^{ab}_{\tau,\tau'} + \sum\limits_i^N \chi_i^a(\tau)\chi_i^b(\tau')\Big) =
\int \mathcal{D}G\, \mathcal{D} \Sigma\,\, e^{ {1\over 2}\sum\limits^n_{a,b} \int d\tau d\tau'    
\Sigma^{ba}_{\tau',\tau}
\Big(N G^{ab}_{\tau,\tau'} + \sum\limits_i^N \chi_i^a(\tau)\chi_i^b(\tau')\Big)}
\end{equation}
The Majoranas are now decoupled from each other 
and may be integrated out leading to the Pfaffian
$\left|\partial_\tau \delta^{ab} + \Sigma^{ab}_{\tau,\tau'}\right|^{N\over 2}$. 
As a result, $N$ enters the action  as a large parameter,
%\begin{equation}
%Z=\int \mathcal{D}G\, \mathcal{D} \Sigma \mathcal{D} \chi \,\, e^{-S[\Sigma,G,\chi]},
%\end{equation}
\begin{equation}
                        \label{eq:action}
- S[\Sigma,G]={N\over 2}\left[ \mathrm{Tr}\log(\partial_{\tau}\delta^{ab} + \Sigma^{ab}_{\tau,\tau'})  +  
 \frac{J^2}{4} \big[G^{ab}_{\tau,\tau'}\big]^4 +  \Sigma^{ba}_{\tau',\tau} G^{ab}_{\tau,\tau'}\right]. 
\end{equation}
This is justification to focus on saddle point configurations which are described by the self-consistent 
Dyson equations~\cite{Sachdev:2015}
\begin{eqnarray}
              \label{eq:Dyson}
 - (\partial_{\tau} + \Sigma)\cdot G = 1; \quad \quad \Sigma=J^2 \big[G\big]^3.
\end{eqnarray}
The first equation here is a matrix equation while in the second the cube operation acts on each matrix element separately, i.e. $[G]^3\equiv \{G_{ \tau, \tau'}^3\}$.  These equations may be solved in the long time limit $\tau-\tau'\gg 1/J$, where $\partial_{\tau}$ may be neglected. The resulting replica diagonal solution acquires  the scale-invariant form:
\begin{equation}
               \label{eq:bare-solution}
G^{ab}_{\tau-\tau'}= - \frac{b}{J^{1/2}}\,\frac{\delta^{ab}\,\mathrm{sgn}(\tau-\tau')}{|\tau-\tau'|^{1/2}}; \quad \quad
\Sigma^{ab}_{\tau-\tau'} =  -  b^3J^{1/2}\,  \frac{\delta^{ab}\,\mathrm{sgn}(\tau-\tau')}{|\tau-\tau'|^{3/2}},
\end{equation}
where $b=(4\pi)^{-1/4}$, or 
\begin{equation}
\label{eq:bare_GS_e}
G^{ab}_\epsilon = - i \delta^{ab}\sqrt{2\pi}\, b\, \mathrm{sgn}(\epsilon)\, \bigl | \epsilon J\bigr |^{-1/2}; \qquad
\Sigma^{ab}_\epsilon = - i \delta^{ab} \sqrt{8\pi}\, b^3\, \mathrm{sgn}(\epsilon)\, \bigl | \epsilon J \bigr |^{1/2}
\end{equation}
in the energy representation (we work at $T \to 0$.) The proportionality $\Sigma\sim
|\epsilon J|^{1/2}$ indeed justifies omission of  $\partial_\tau\to -i\epsilon$ in the trace of
the logarithm, as long as $\epsilon\ll J$.

We next turn to the discussion of fluctuations around the above mean field configuration. As with other models containing quenched randomness, one might expect soft fluctuations away from the replica-diagonal configurations $\propto \delta^{ab}$. However such fluctuations appear to play a lesser role, an observation corroborated by the absence of significant mesoscopic fluctuations in the model. Indeed, our numerics shows that correlation functions in any single realization are essentially self-averaging which suggests that  the replica index does not play a role, at least for energies larger than the inverse of the many body level spacing, $J\epsilon \gtrsim 2^{-N/2}$. We will ignore the replica structure of the theory throughout.

However, there exists a second class of fluctuations which is of high relevance to the description of correlation functions for energies in the much larger range $2^{-N/2}\lesssim J\epsilon \lesssim 1/N\ln N$. As has been realized in Refs.~\cite{Kitaev:2015,Polchinski:2016,Maldacena:2016}, Eq.~\eqref{eq:bare-solution} is but one representative of a whole manifold of solutions of Eqs.~(\ref{eq:Dyson}) in the limit  $\partial_\tau \to 0$. This manifold  is obtained by reparameterizations of time as $\tau \to f(\tau)$ where $f(\tau)$ is an arbitrary {\em monotonic} differentiable function. 
It is straightforward to verify that the $\partial_\tau\to 0$  saddle point equations  (\ref{eq:Dyson}) are invariant under the  transformation
\begin{eqnarray}
               \label{eq:soft-manifold}
G_{\tau,\tau'} &=&  - \frac{b}{J^{1/2}}\mathrm{sgn}(f(\tau)-f(\tau'))\,\frac{f'(\tau)^{1/4} f'(\tau')^{1/4}}{|f(\tau)-f(\tau')|^{1/2}}, \\
\Sigma_{\tau,\tau'}  &=&  -  b^3J^{1/2}\mathrm{sgn}(f(\tau)-f(\tau'))\,  \frac{f'(\tau)^{3/4} f'(\tau')^{3/4} }{|f(\tau)-f(\tau')|^{3/2}}.
\label{eq:soft-manifold-1}
\end{eqnarray}
Reparameterization transformations $f(\tau)$ are generated by the infinite dimensional Virasoro algebra. One may verify that the saddle points remain invariant under the finite dimensional group of conformal transformations $\mathrm{SL}(2,R)$, i.e. in effect we are met with an infinite dimensional Goldstone mode manifold of reparameterizations modulo the unbroken $\mathrm{SL}(2,R)$-transformations.

%{\bf Global group $SL(2,R)$}.

This observation dictates the subsequent strategy for the computation of  correlation functions in the long time limit: (i) one has to perform a Gaussian integration  over massive fluctuations of the fields $G$ and $\Sigma$, i.e. modes different from those described by Eqs.~(\ref{eq:soft-manifold}), (\ref{eq:soft-manifold-1}). This was essentially accomplished in Ref.~\cite{Maldacena:2016}. (ii)
The resulting expressions should be reparametrized with $\tau \to f(\tau)$ and the rules of Eqs.~(\ref{eq:soft-manifold}),  (\ref{eq:soft-manifold-1}) and (iii) averaged over all realizations of $f(\tau)$. Since these latter fluctuations are soft, the corresponding functional integration should be performed exactly rather than in a Gaussian  approximation.  We will show that this program leads to the emergence of  Liouville quantum mechanics in the picture and in the consequence to  universal $\propto \tau^{-3/2}$ scaling of all long-time correlation functions.

To perform the crucial third step of the program we need to know the effective theory governing the reparameterization modes. An effective action for the $f$-modes was  recently proposed by Maldacena,  Stanford, and  Yang~\cite{Maldacena:2016, Maldacena:2016a}. In the next section we extend this construction by deriving a generalized form of the action and the corresponding integration \emph{measure}, and by introducing a variable transform which will be instrumental to all further calculations.  
In Sec.~\ref{sec:G2} we apply the theory to the non-perturbative calculation of the two-point Green function and in Sec.~\ref{sec:G4} to that of the  four-point function.  

\section{Soft mode action}
\label{sec:Soft_Mode_Action}

In the conformal limit, where one neglects $\partial_\tau$, the action is invariant under the transformation (\ref{eq:soft-manifold}). A finite action cost of the $f(\tau)$-fluctuations  follows from an expansion of $\mathrm{Tr}[\log(\partial_{\tau} + \Sigma)-\log(\Sigma)]=\mathrm{Tr}\log(1- \partial_{\tau} G)$   in powers of $\partial_\tau$, where the leading approximation $\Sigma^{-1}= - G$ has been used.   The first non-vanishing term is of second order in derivatives, and one obtains
\begin{equation}
            \label{eq:expansion}
S[f]= {N\over 4}\, \mathrm{Tr} (\partial_{\tau} G\partial_{\tau} G)=- \frac{b^2N}{16J} \int\!\!\!\!\int d\tau d\tau' \,
\frac{f'(\tau)^{3/2} f'(\tau')^{3/2}}{|f(\tau)-f(\tau')|^{3}}\,. 
\end{equation}
This action is inherently non-local. As we show in~\ref{app:Action} one may pass   to the variable
 $t=f(\tau)$ and performing a Fourier transformation to obtain a  non-analytic kernel of the form $\omega^2\log(J/|\omega|)$. Its logarithmic dependence is a consequence of the slow decay $G_{\tau,\tau'} G_{\tau',\tau}\propto |\tau-\tau'|^{-1}$  of the bare correlation functions at large time, cf. Eq.~(\ref{eq:bare-solution}). As we shall see below, the actual two-point propagator decays as $|\tau-\tau'|^{-3/2}$ beyond a certain time scale. This means that  the logarithm is to be cut at a frequency scale, $\Delta$, which will be self-consistently  determined below. The kernel regularized in this way is analytic, $\sim \omega^2 \log(J/\Delta)$, and is local in time. The corresponding soft-mode action 
\begin{equation}
            \label{eq:soft-action}
S[f]={M\over 2} \int d\tau \left(\frac{f''(\tau)}{ f'(\tau)}\right)^2\,, \quad\quad\quad M=\frac{b^2}{32 J}\,N\log\left({J\over\Delta}\right), 
\end{equation}  
has recently been suggested by Maldacena \& Stanford \cite{Maldacena:2016}, albeit possibly with a different coefficient $M$. Emphasizing its unbroken $\mathrm{SL}(2,R)$-symmetry they wrote it in the form of a
Schwarzian derivative\footnote{The Schwarzian derivative is defined as $\{f,\tau\} \equiv \frac{f'''}{f'} - \frac 32 \left( \frac{f"}{f'}\right)^2 \equiv   \left( \frac{f''}{f'}\right)'- \frac 12 \left( \frac{f"}{f'}\right)^2$.  }, 
$S \propto - \int d\tau \{f,\tau\}$. However, one should keep in mind that the actual $\mathrm{SL}(2,R)$-invariant soft mode action~Eq.~(\ref{eq:expansion}) is intrinsically non-local. The distinction becomes  more substantial for models with generic $q$-body interactions~\cite{Maldacena:2016} (in our case
$q=4$) for which the denominator in Eq.~(\ref{eq:expansion}) acquires the form $|f(\tau)-f(\tau')|^{2+4/q}$, or  $\sim |\omega|^{1+4/q}$. This kernel is essentially non-local  and it is not clear if it can be brought into a time-local form. In praticular, for $q\to \infty$~\cite{Maldacena:2016} one obtains the purely dissipative Caldeira-Leggett kernel $|\omega|$ which at sufficiently strong coupling~\cite{Schmid,Leggett}  completely suppresses all  quantum coherence. However, for the model presently under consideration, i.e. $q=4$, the local approximation to the action seems appropriate. 

The specific dependence of the action on ratios of time derivatives of $f$ suggests the variable transform
\begin{equation}
\label{eq:phi}
f'(\tau)=e^{\phi(\tau)}; \quad\quad\quad   f(\tau)-f(\tau') = \int\limits_{\tau'}^{\tau} \!\!d\tau\, e^{\phi(\tau)},
\end{equation}
which is well defined due to the assumed temporal monoticity of $f(\tau)$. 
Expressed in terms of the $\phi$-variables, the  soft mode action becomes quadratic,
\begin{equation}
            \label{eq:phi-action}
S[\phi]={M\over 2} \int \!\! d\tau\, \big[\phi'(\tau)\big]^2. 
\end{equation}  
We observe that one of the basic $\mathrm{SL}(2,R)$-transformations -- rescaling of time, $f(\tau)\to \alpha f(\tau)$ -- translates to a shift of the $\phi$-field, $\phi\to \phi +\ln\alpha$. 
The action (\ref{eq:phi-action}) is manifestly invariant under this transformation. 

The integration over soft modes amounts to a functional integral over $f(\tau)$ or equivalently over $\phi(\tau)$ with effective action Eqs.~(\ref{eq:soft-action}) or (\ref{eq:phi-action}), respectively. To complete the description of the procedure we need to specify the corresponding integration \emph{measures}. We show in~\ref{app:Measure} that in the $\phi$--representation the measure is {\em flat} and therefore this choice of variables is particularly convenient.

\section{Two-point Green function}
\label{sec:G2}

We are now in a position to average the two-point Green function (\ref{eq:soft-manifold}) over soft mode fluctuations. To this end we substitute the reparameterization Eq.~\eqref{eq:phi} in  Eq.~(\ref{eq:soft-manifold}) and integrate  over $\phi$  with the weight (\ref{eq:phi-action}) and the flat measure.  This leads to 
\begin{eqnarray}
G(\tau-\tau') &=& \mp \frac{b}{\sqrt{J}} \int\!\! {\cal D}\phi\,\, \frac{e^{{1\over 4}\phi(\tau)} e^{{1\over 4}\phi(\tau')}}{\left[ \int\limits_{\tau'}^{\tau} \!\!d\tau\, e^{\phi(\tau)}\right]^{1/2}}\,\, e^{-{M\over 2}\int d\tau [\phi']^2} \nonumber \\
\label{eq:two-point}
&=& \mp \frac{b}{\sqrt{\pi J} }\int\limits_0^\infty \!\! \frac{d\alpha}{\sqrt{\alpha} } \int\!\! {\cal D}\phi\, e^{{1\over 4}\phi(\tau)} e^{{1\over 4}\phi(\tau')}   \,\, e^{-{M\over 2}\int d\tau [\phi']^2 - \alpha \int\limits_{\tau'}^{\tau} \!\!d\tau\, e^{\phi(\tau)}},  
\end{eqnarray}
where we used the identity 
\begin{equation}
\label{eq:identity} 
{\cal A}^{-p}=\frac{1}{\Gamma(p)}  \int\limits_0^\infty \!\! d\alpha\, \alpha^{p-1}\, e^{-\alpha {\cal A}}. 
\end{equation}
and the global sign accounts for the factor $\mathrm{sgn}(\tau-\tau')$. 

Interpreting this expression as an (imaginary time) path integral,  the Green function becomes an expectation value  $\langle e^{{1\over 4}\phi(\tau)} e^{{1\over 4}\phi(\tau')}\rangle$ taken in a time dependent variant  of  Euclidean quantum mechanics, i.e. the quantum mechanics of a point particle with $\alpha e^\phi$ potential. The problem is time-dependent in that the potential is switched on only during the time window $[\tau', \tau]$, i.e. we are dealing with some kind of quantum quench. We note that previous applications of
Liouville quantum mechanics to random systems include the description of Dirac fermions in a disorder potential \cite{Balents:1997, Kogan:1996}. Incidentally, it may also be considered as the one-dimensional limit of two-dimensional Liouville field theory, where the latter plays an important role in two-dimensional holographic contexts~\cite{Zamolodchikov:1996}. 

Switching from the path integral language to a spectral decomposition in terms of quantum mechanical wave functions we obtain 
\begin{equation}
\label{eq:quantum-mechanics}
G(\tau-\tau') = \mp  \frac{b}{\sqrt{\pi J} }  \int\limits_0^\infty \!\! \frac{d\alpha}{\sqrt{\alpha} }  \sum\limits_k 
\langle 0| e^{{1\over 4}\phi}|k\rangle \, e^{-E_k|\tau-\tau'|}\, \langle k| e^{{1\over 4}\phi}|0\rangle,
\end{equation}
where $|0\rangle = \mathrm{const}$ is the groundstate before (or after) the Liouville potential $ \alpha e^\phi$ is switched on (or off), and 
$\langle \phi|k\rangle= \Psi_k(\phi)$ is an eigenfunction of the corresponding Schr\"odinger equation
\begin{equation}
\label{eq:Schrodinger}
-{1\over 2M}\, \partial^2_\phi \Psi_k(\phi) +\alpha e^\phi \Psi_k(\phi)  = E_k\Psi_k(\phi),
\end{equation}
with the eigenvalue $E_k$. It has a continuous spectrum labeled by a quantum number $k>0$, such that 
$E_k =k^2/2M$ and 
\begin{equation}
  \label{eq:eigenfunction}
\Psi_k(\phi)={\cal N}_k K_{2ik}\left(2\sqrt{2 M\alpha}\, e^{\phi/2} \right), \quad\quad\quad {\cal N}_k= \frac{2}{\Gamma( 2 i k)},
\end{equation}
where $K_{2ik}(x)$ is a modified Bessel function. The normalization factor ${\cal
N}_k$ is chosen such that  at $\phi\to-\infty$ the state becomes a standing wave normalized
according to $\langle k| k' \rangle = 2\pi\, \delta(k-k')$. Using these results, the matrix elements in~\eqref{eq:quantum-mechanics} become
\begin{eqnarray}
\langle 0| e^{{1\over 4}\phi}|k\rangle&=& {\cal N}_k \int\limits_{-\infty}^\infty\!\! d\phi\, e^{{1\over 4}\phi}\,\, K_{2ik}\left(2 e^{(\phi+\ln(2\alpha M))/2} \right)= \frac{{\cal N}_k}{(2 M \alpha)^{1/4}} \int\limits_{-\infty}^\infty\!\! d\phi\, e^{{1\over 4}\phi}\,\, K_{2ik}\left(2 e^{\phi/2} \right) \nonumber \\
\label{eq:matrix-elements}
&=& \frac{1}{(2 M \alpha)^{1/4}} \,\, \frac{\Gamma(\frac 14 - i k) \Gamma(\frac 14 + i k) }{\Gamma(2 i k)}.
\end{eqnarray}
Notice that $v=\ln\alpha$ represents a constant shift of the $\phi$-field -- one of the global $\mathrm{SL}(2,R)$ transformations -- which actually should be excluded from the slow mode integration. Within the quantum mechanical framework its redundancy shows in that it factors out of the matrix elements and enters as a formally divergent factor $\int_0^\infty d\alpha/\alpha= \int dv$. Much as in a theory with gauge symmetries, this formally infinite constant should be ignored, or equivalently $\alpha$ locked by a `gauge fixing convention'. This way we arrive at 
\begin{equation}
\label{eq:G-semi-final}
G(\tau)  = - \mathrm{sgn}(\tau) \frac{b}{\sqrt{J}} \left(\frac{2}{\pi M} \right)^{1/2} \int\limits_0^{\infty}\!\! dk\,\, e^{-k^2|\tau|/2M} \, {\cal M}_2(k), % \quad \quad  {\cal M}_2(k)=....
\end{equation}
where the spectral density takes the form
\begin{equation}
 {\cal M}_2(k) = \frac{k \sinh(2\pi k)}{2\pi^2}  \,\, \Gamma^2\left(\tfrac{1}{4} + i k\right)  \Gamma^2\left(\tfrac{1}{4} - i k\right). 
 \end{equation}
The latter is normalized  such that ${\cal M}_2(k) \to 1$ for $k\gg 1$. For short
times $|\tau|\ll M$, where large values of $k$ dominantly contribute to the integral,
we therefore  obtain $G(\tau)\sim |\tau|^{-1/2}$, independent of  $M$. This is
 but the familiar scale-invariant  Green function,
Eq.~(\ref{eq:bare-solution}), with the same numerical prefactor. In the opposite
limit, $k\ll 1$, one finds ${\cal M}_2(k)\propto k^2$, leading to $G(\tau)\propto
M/|\tau|^{3/2}$ at $\tau\gg M$. In the energy representation this reads
$G(\epsilon)\propto M\sqrt{|\epsilon|}$ for $\epsilon\ll M^{-1}$. As a result the
$|\epsilon|^{-1/2}$ scaling of the single-particle density of states (DoS) at large
energies crosses over to a $|\epsilon|^{1/2}$ suppression at small energy. This
behavior is reminiscent to the so-called zero-bias anomaly  known in the theory of
interacting fermions in presence of the random potential~\cite{AA1, Nazarov-book,
Kamenev:1999}. It may also be understood as a manifestation of the Mermin-Wagner theorem which requires that the breaking of a continuous (reparameterization) symmetry by the Green function mean field amplitude be reversed in the limit where the explicit symmetry breaking ($\partial_t\sim \epsilon$) vanishes.

\begin{figure}[t]
\includegraphics[width=6.6cm]{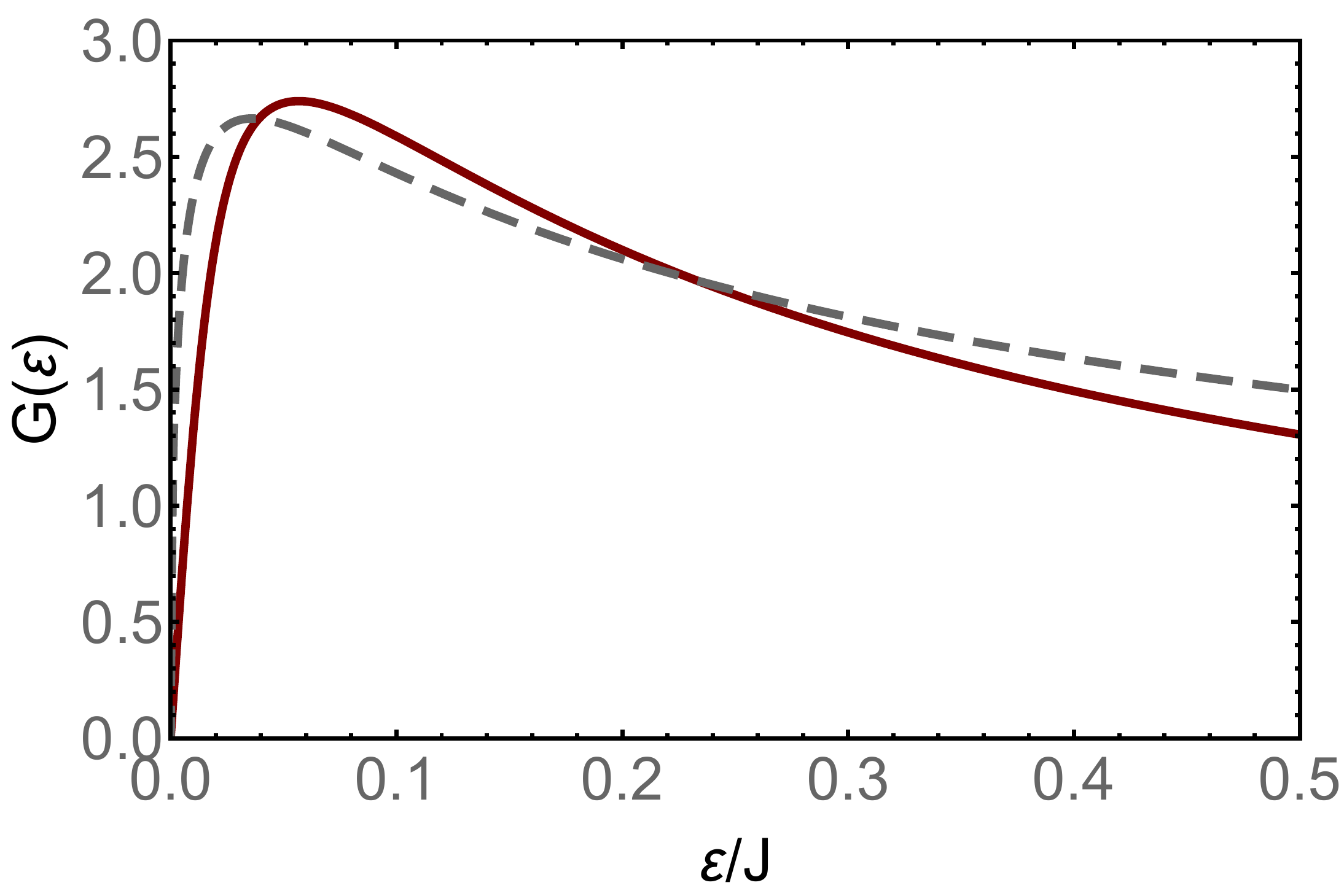}
\hspace{0.5cm}
\includegraphics[width=6.5cm]{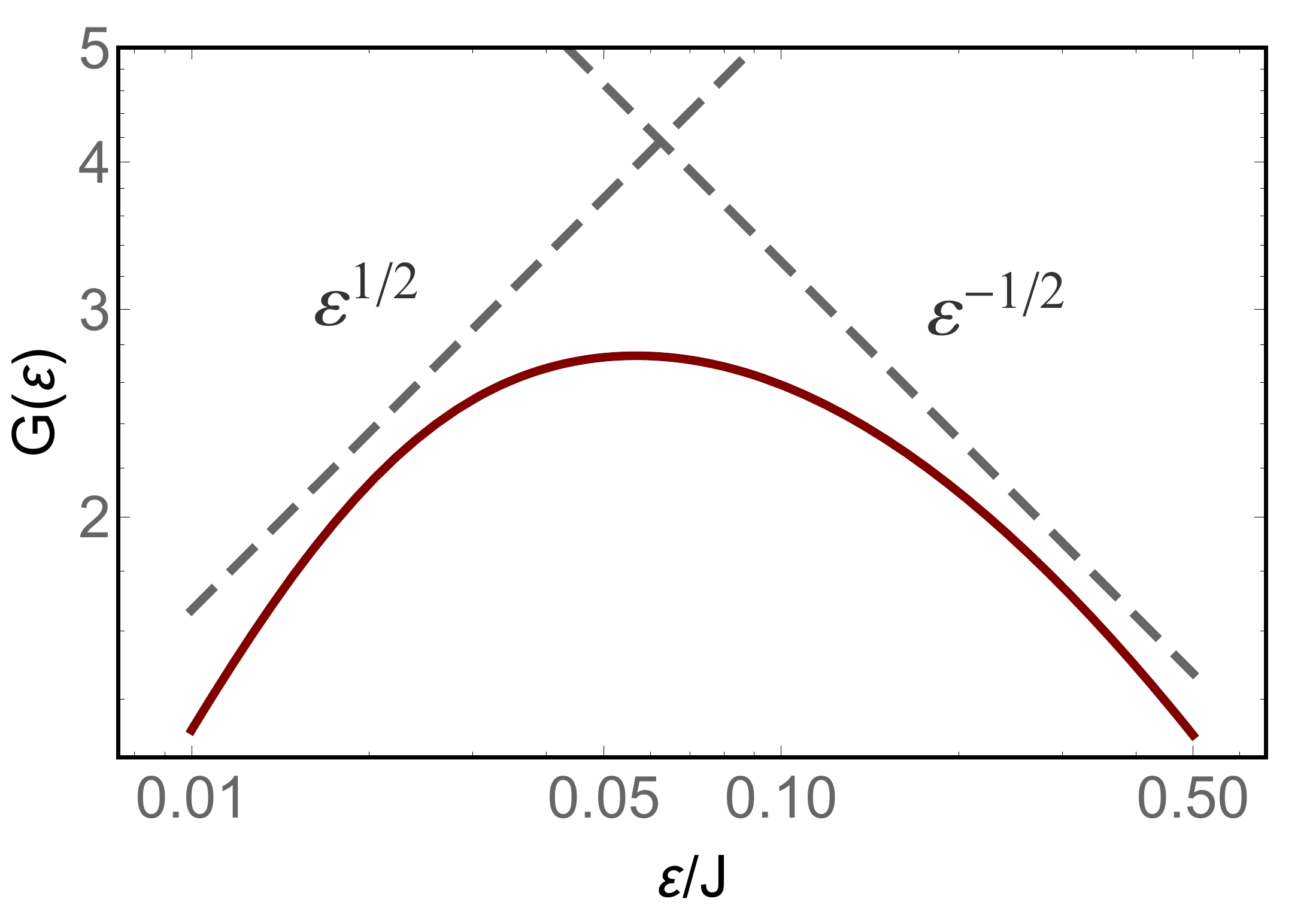}
\caption{ Numerical (solid red line) versus analytic (dashed line) result for the Matsubara Green function $G(\epsilon)$ evaluated for $N=24$ Majorana fermions (left panel).
Logarithmic plot of the numerically evaluated Matsubara Green function shown versus two expected scaling laws (right panel).
}
\label{fig:G}
\end{figure}

The energy dependence of the two-point Green function in the crossover region can be obtained from the full spectral representation
\begin{equation}
G(\epsilon) =  - \frac{i b}{\sqrt{J}} \left(\frac{2}{\pi M} \right)^{1/2} \int_0^{+\infty}\!\! dk\, {\cal M}_2(k) \,\frac{ 2 \epsilon}{ E_k^2 + \epsilon^2}, \quad E_k = k^2/2M.
\end{equation}
%which upon  analytic continuation can be used to extract the single particle DoS, $\nu(E) = - \frac{1}{\pi} \, \mathrm{Im}\, G( - i E + 0^+)$.  
The result is shown in Fig.~\ref{fig:G} together with the Matsubara Green function of an $N=24$ system obtained from the Lehmann representation by exact diagonalization. 
The latter was constructed for a fixed realization of the disorder and the apparent absence of stochastic fluctuations underpins the irrelevancy of sample-to-sample fluctuations. For small energies, the Green function vanishes, although the agreement with the analytical scaling $\epsilon^{1/2}$ is not perfect. The reason is that the behavior of the Green function at the smallest energies is influenced by the finite {\em many-body} level spacing  whose resolution is beyond the scope of the present theory. For $N=24$, the energy range between the many-body level spacing, $\sim e^{-N/2}$, and the one-body energy scale, $\sim M^{-1}\approx 1/N\log N $, is not wide enough to clearly resolve the predicted  exponent  $\epsilon^{1/2}$. 

\section{Four-point function}
\label{sec:G4}

We turn now to the four-point function defined as 
\begin{equation}
\label{eq:four-point}
N^{-2}\sum\limits_{i,j}^N \langle \,\chi_i(\tau_1)\chi_i(\tau_2)\chi_j(\tau_3)\chi_j(\tau_4)\rangle =\langle G_{\tau_1,\tau_2} G_{\tau_3,\tau_4} \rangle,
\end{equation}
where the angular brackets denote quantum averaging over both massive and soft fluctuations of the $G$ and $\Sigma$ field. The results obtained for this function sensitively depend on the chronological order of the time arguments. It turns out that for $\tau_{2,3}<\tau_{1,4}$ it is sufficient to disregard  the massive  modes, which generally lead to contributions of $\mathcal{O}(1/N)$. To start with, we consider temporal arguments of this ordering and neglect the massive modes.  This will be sufficient to calculate, e.g.,  the polarization operator $\langle G_{\tau,\tau'} G_{\tau',\tau} \rangle$, which enters the soft mode action 
(\ref{eq:expansion}), and thus determines $M$. The complementary regime $\tau_{1,2}<\tau_{3,4}$ in which massive modes play a crucial role will be considered later.

% By contrast, for $\tau_{1,2} <\tau_{3,4}$ the soft mode averaging leads to the
% disconnected part $\langle G_{\tau_1,\tau_2} G_{\tau_3,\tau_4} \rangle=\langle
% G_{\tau_1,\tau_2}\rangle \langle  G_{\tau_3,\tau_4} \rangle$ only. To find any
% connected piece in this regime one must resort to massive modes.

The soft modes dress the bare Green functions $G_{\tau_1,\tau_2}$ and  $G_{\tau_3,\tau_4}$ according to Eq.~(\ref{eq:soft-manifold}). Once more we employ Eq.~(\ref{eq:identity}) to elevate the denominators to an effective action and find  
\begin{equation}
\label{eq:four-point-action}
\langle G_{\tau_1,\tau_2} G_{\tau_3,\tau_4} \rangle \propto 
\int\!\!\!\!\int\limits_0^\infty \!\! \frac{d\alpha d\beta}{\sqrt{ \alpha\beta} } \int\!\! {\cal D}\phi\, e^{{1\over 4}\phi(\tau_1)} e^{{1\over 4}\phi(\tau_2)}   e^{{1\over 4}\phi(\tau_3)} e^{{1\over 4}\phi(\tau_4)} \,\, e^{-{M\over 2}\int d\tau [\phi']^2 - \alpha \int\limits_{\tau_2}^{\tau_1} \!\!d\tau\, e^{\phi(\tau)} 
- \beta \int\limits_{\tau_3}^{\tau_4} \!\!d\tau\, e^{\phi(\tau)}},  
\end{equation}
where we assumed $\tau_1 > \tau_2$ \& $\tau_3 < \tau_4$.
This is again an expectation value of four time ordered operators $e^{{1\over 4}\phi}$ taken in the  quantum ground state of a system  with time-dependent potential $\alpha e^{\phi}$ for $\tau \in [\tau_2,\tau_1]$ and $\beta e^{\phi}$ for $\tau \in [\tau_3,\tau_4]$. All subsequent calculations crucially depend on the ordering of the four times. We consider two distinct configurations: 

\subsection{Time ordering: $\tau_{2},\tau_{3} < \tau_{1},\tau_{4}$}
To be specific, let us first take the ordering $\tau_2<\tau_3<\tau_1<\tau_4$. The corresponding four-point function (\ref{eq:four-point}) averaged over the soft modes acquires the form    
\begin{eqnarray}
\langle G_{\tau_1,\tau_2} G_{\tau_3,\tau_4} \rangle &\propto&
\int\!\!\!\!\int\limits_0^{\infty} \!\! \frac{d\alpha d\beta}{\sqrt{ \alpha\beta} }   
\sum\limits_{k_\alpha,k_\beta,k_{\alpha+\beta}} 
\langle 0| e^{{1\over 4}\phi}|k_\alpha \rangle \, e^{-\frac{k_\alpha^2}{2M} (\tau_3-\tau_2)} 
\langle k_\alpha| e^{{1\over 4}\phi}|k_{\alpha+\beta}\rangle e^{-\frac{k_{\alpha+\beta}^2}{2M} (\tau_1-\tau_3)} \nonumber \\
\label{eq:quantum-mechanics-four-point}
&\times&\langle k_{\alpha+\beta}| e^{{1\over 4}\phi}|k_{\beta}\rangle e^{-\frac{k_{\beta}^2}{2M} (\tau_4-\tau_1)}\langle k_{\beta}| e^{{1\over 4}\phi}|0\rangle,
\end{eqnarray} 
where the state
\begin{equation}
  \label{eq:eigenfunction-1}
|k_\sigma \rangle = \Psi_{k_\sigma}(\phi)={\cal N}_{k_\sigma} K_{2ik_\sigma}\left(2\sqrt{ 2 M\sigma}\, e^{\phi/2} \right)
\end{equation}
with $\sigma=\alpha,\beta,(\alpha+\beta)$. Computing the matrix elements as integrals in the $\phi$-representation and applying a shift of integration variables    $\phi\to \phi +\ln\alpha$ and $\phi\to \phi + \ln\beta$ in the first two and last two, respectively, one notices that the
matrix elements depend only $u=\ln(\alpha/\beta)$. The complementary variable $v=\ln(\alpha\beta)$ drops out and enters as a multiplicative and formally divergent integral $\int d v$.
This is again a manifestation of the unbroken $\mathrm{SL}(2,R)$ invariance, i.e. the integral should be ignored.  
All remaining integrals, including $\int d u$, are convergent and may be, in principle, evaluated. The resulting expressions are tedious and not really illuminating. 

The situation simplifies if we set  $\tau_2=0^-; \tau_3=0^+$ and $\tau_1=\tau^-; \tau_4=\tau^+$ to effectively consider a time ordered correlation function $\sim \langle (\chi_i\chi_j)(0)\,(\chi_i\chi_j)(
\tau)\rangle $. In this case the corresponding propagators, e.g. $\sum_{k_\sigma} |k_\sigma \rangle \langle 
k_\sigma|=1$, where $\sigma=\alpha,\beta$, and the  four-point function  simplifies to
\begin{eqnarray}
\langle G_{\tau,0} G_{0,\tau} \rangle &\propto&  
\int\!\!\!\!\int\limits_0^\infty \!\! \frac{d\alpha d\beta}{\sqrt{ \alpha\beta} }   
\sum\limits_{k_{\alpha+\beta}}  
\langle 0|  e^{{1\over 2}\phi}|k_{\alpha+\beta}\rangle e^{-\frac{k_{\alpha+\beta}^2}{2M} \tau}\langle k_{\alpha+\beta}| e^{{1\over 2}\phi} |0\rangle \nonumber\\
\label{eq:polarization}
&=&\frac{1}{2M}\iint\limits_0^\infty \!\! \frac{d\alpha d\beta}{\sqrt{ \alpha\beta} }\, \frac{1}{\alpha+\beta}   
\sum\limits_{\tilde k}  
\langle 0|  e^{{1\over 2}\phi}|\tilde k\rangle e^{-\frac{k^2}{2M} \tau}\langle \tilde k| e^{{1\over 2}\phi} |0\rangle,
\end{eqnarray} 
where in the last expression we performed the shift $\phi\to\phi +\ln(\alpha+\beta)$ so that the remaining matrix elements do not contain $\alpha$ and $\beta$ and $| \tilde k \rangle \equiv  {\cal N}_k K_{2ik} (2 e^{\phi/2})$.
As mentioned above 
\begin{equation}
\int\!\!\!\!\int\limits_0^\infty \!\! \frac{d\alpha d\beta}{\sqrt{ \alpha\beta} }\, \frac{1}{\alpha+\beta}   =
{1\over 4}\int\!\!\!\!\int\limits_{-\infty}^\infty \!\! \frac{du dv}{\cosh (u/2)}= {\pi\over 2}\int dv \, ,
\end{equation}
where $u=\ln(\alpha/\beta)$ and $v=\ln(\alpha\beta)$, is a spurious factor due to the
redundant integration over a global rescaling of time. The remaining sum
$\sum_{\tilde k}\ldots$ in Eq.~(\ref{eq:polarization}) is the Liouville quantum
mechanics expectation value of $\langle e^{{1\over 2}\phi(0)}e^{{1\over
2}\phi(\tau)}\rangle$. Comparing it to Eq.~(\ref{eq:quantum-mechanics}), one notices
that the only difference to the two-point function is that now each of the end points
contains two factors $e^{{1\over 4}\phi}$, instead of one. The subsequent calculations
are essentially similar to that following  Eq.~(\ref{eq:quantum-mechanics}),
resulting in
\begin{equation}
\label{eq:polarization-semi-final}
\langle G_{\tau,0} G_{0,\tau} \rangle =\frac{b^2}{J M}\int_0^{+\infty}\!\! dk\, e^{-k^2|\tau|/2M} \, {\cal M}_4(k),\quad \quad  {\cal M}_4(k)= k \tanh(\pi k).
\end{equation} 
For $k\gg 1$, ${\cal M}_4(k)\propto k$ and thus $\langle G_{\tau,0} G_{0,\tau}
\rangle\propto |\tau|^{-1}$  is $M$-independent. This is nothing but the disconnected
part of the polarization operator, $ \langle G_{\tau,0} G_{0,\tau} \rangle = \langle
G_{\tau,0}\rangle \langle G_{0,\tau} \rangle$, valid for $\tau\ll M$, which allows to
restore proper normalization in Eq.~(\ref{eq:polarization-semi-final}). This
factorization breaks down at $\tau>M$. Indeed, we saw in the previous section that at
long times $G^2(\tau)\propto |\tau|^{-3}$. However the polarization operator
(\ref{eq:polarization-semi-final})  exhibits much slower decay. Noticing that ${\cal
M}_4(k)\propto k^2$  at $k\ll 1$, one finds $\langle G_{\tau,0} G_{0,\tau}
\rangle\propto M^{1/2} |\tau|^{-3/2}$ for $\tau\gg M$.
 
The structure of the expressions above implies that any $2p$-point correlation function with $2p$ times grouped into two clusters separated by a time $\tau$ reduces to
 \begin{equation}
\label{eq:2p}
\langle G_{\tau,0}^p \rangle = \frac{b^p}{J^{p/2}}  \left\langle e^{{p\over 4}\phi(0)}e^{{p\over 4}\phi(\tau)}\right\rangle = \frac{b^p}{(J M)^{p/2}} \times \frac{2^{1-p/2}}{\Gamma(p/2)}
 \int_0^{+\infty}\!\! dk\, e^{-k^2|\tau|/2M} \, {\cal M}_{2p}(k), %\qquad  {\cal M}_{2p}(k)=\frac{1}{4\pi}\left|\langle 0|e^{{p\over 4}\phi}|k\rangle\right|^2
\end{equation}  
where the spectral function 
\begin{equation}
{\cal M}_{2p}(k) = \frac{1}{4\pi}\left|\langle 0|e^{{p\over 4}\phi}|\tilde k\rangle\right|^2 =  \frac{k \sinh(2\pi k)}{2\pi^2} \, \, \Gamma^2\left(\tfrac{p}{4} + i k\right)  \Gamma^2\left(\tfrac{p}{4} - i k\right),
\end{equation}
 generalizes our previous results for $p=1,2$. 
We observe that if $k\gg 1$ then ${\cal M}_{2p}(k)\propto k^{p-1}$,  and thus $\langle G_{\tau,0}^p \rangle=\langle G_{\tau,0} \rangle^p\propto |\tau|^{-p/2}$. On the other hand, the small $k\ll 1$ behavior of the matrix elements is universal ${\cal M}_{2p}(k)\propto k^{2}$. As a result, we arrive at the universal long time decay of the connected correlation functions: 
 \begin{equation}
\label{eq:universal}
\langle G_{\tau,0}^p \rangle \propto  \frac{M^{(3-p)/2}}{|\tau|^{3/2}}\,.
\end{equation}  
Similar universality was found \cite{Shelton:1998,Balents:1997} within the context of
Dirac fermions in a random potential. There, the wave-function moments: $\langle
\sum_n |\psi_n(0)\psi_n(x)|^p\rangle$ was found to decay as  $x^{-3/2}$ independent
on the index $p$. This behavior was attributed to the influence of tails of the
probability distribution functions, which emphasize rare atypical realizations of the
disorder, on the moments of wave functions. In our case the correlations functions
are self-averaging vis-a-vis disorder realizations. It is thus large quantum
fluctuations, described by soft reparametrization modes, which lead to the universal
behavior~(\ref{eq:universal}).

We are now in a position to fix the constant $M$ in the soft-mode action (\ref{eq:soft-action}). The logarithmic singularity of the kernel at small frequency originates in the slow decay of the 
{\em bare}  polarization operator $ \langle G_{\tau,0}\rangle \langle G_{0,\tau} \rangle\propto 1/|\tau|$. 
However, if one uses the full connected polarization operator $\langle G_{\tau,0} G_{0,\tau} \rangle$ to derive the action, the logarithmic singularity is cut of at $\tau\approx M$, beyond which it decays as $|\tau|^{-3/2}$. As a result, the constant in the action (\ref{eq:soft-action}) is determined by the self-consistent equation
\begin{equation}
\label{eq:self-consistent}
M=\frac{b^2 }{32 J}\,N\log(J M)\,,\qquad M\approx \frac{b^2}{32 J} \, N\log N ,
\end{equation}
which we resolved here with  logarithmic accuracy. Hence one finds that the effective single particle mean-level spacing $\Delta$ at small energy is reduced in comparison with the naive expectation $J/N$ and is actually given by $\Delta\sim J/(N\log N)$.   This is due to the fact that the bare single-particle DoS is enhanced at small energy.   

\subsection{Time ordering: $\tau_{1},\tau_{2} < \tau_{3},\tau_{4}$}     

Finally we briefly comment on an alternative time-ordering arrangement in the correlation function. In this case, the massive modes (completely disregarded above) are of vital importance to the description of the  
connected part of the correlation function. We begin by showing that at the above time configuration the pure soft mode action incorrectly predicts a decoupling of  the  four-point correlation function (\ref{eq:four-point}) into a product of two two-point functions. The underlying reason is that the Liouville potentials 
$\alpha e^\phi$ and $\beta e^{\phi}$ are switched on during the time windows $\tau\in [\tau_1,\tau_2]$ and 
$\tau\in [\tau_3,\tau_4]$ correspondingly, while for $\tau\in [\tau_2, \tau_3]$ they are off. The eigenstates in that time window are simple plane waves $|q\rangle =e^{iq\phi}$ and this leads to the expression, cf.~Eq.~(\ref{eq:quantum-mechanics-four-point}), 
\begin{eqnarray}
 \langle G_{\tau_1,\tau_2} G_{\tau_3,\tau_4} \rangle &\propto&
\int\!\!\!\!\int\limits_0^\infty \!\! \frac{d\alpha d\beta}{\sqrt{ \alpha\beta} }   
\sum\limits_{k_\alpha,k_\beta,q}  
\langle 0| e^{{1\over 4}\phi}|k_\alpha \rangle \, e^{-\frac{k_\alpha^2}{2M} (\tau_2-\tau_1)} 
\langle k_\alpha| e^{{1\over 4}\phi}|q\rangle e^{-\frac{q^2}{2M} (\tau_3-\tau_2)} \nonumber\\
\label{eq:quantum-mechanics-four-point-1}
&\times&\langle q| e^{{1\over 4}\phi}|k_{\beta}\rangle e^{-\frac{k_{\beta}^2}{2M} (\tau_4-\tau_3)}\langle k_{\beta}| e^{{1\over 4}\phi}|0\rangle,
\end{eqnarray} 
where the wave-functions $|k_{\sigma}\rangle $, $\sigma=\alpha,\beta$  are given by Eq.~(\ref{eq:eigenfunction-1}). Proceeding as before we perform shifts $\phi\to \phi+\ln \sigma$ in the matrix elements to arrive at    
\begin{eqnarray}
\langle G_{\tau_1,\tau_2} G_{\tau_3,\tau_4} \rangle &\propto&
\int\!\!\!\!\int\limits_0^\infty \!\! \frac{d\alpha d\beta}{\alpha\beta }\, e^{-iq(\ln\alpha-\ln\beta)}     
\sum\limits_{k,k',q}  
\langle 0| e^{{1\over 4}\phi}|k \rangle \, e^{-\frac{k^2}{2M} (\tau_2-\tau_1)} 
\langle k| e^{{1\over 4}\phi}|q\rangle e^{-\frac{q^2}{2M} (\tau_3-\tau_2)}\nonumber\\
\label{eq:quantum-mechanics-four-point-2} 
&\times&
\langle q| e^{{1\over 4}\phi}|k'\rangle e^{-\frac{k'^2}{2M} (\tau_4-\tau_3)}\langle k'| e^{{1\over 4}\phi}|0\rangle,
\end{eqnarray} 
where $\alpha,\beta$ appear only in the integral
\begin{equation}
\int\!\!\!\!\int\limits_0^\infty \!\! \frac{d\alpha d\beta}{\alpha\beta }\, e^{-iq(\ln\alpha-\ln\beta)}  = {1\over 2} 
\int\!\!\!\!\int\limits_{-\infty}^\infty \!\! dv du \,e^{-iq u} =\pi \delta(q)\int dv. 
\end{equation}
Here, $v=\ln(\alpha\beta)$ and $u=\ln(\alpha/\beta)$, as before, and the spurious $\int dv$ integral should  again be omitted. However, the emergence of the factor $\delta(q)$  is  new and remarkable. It enforces that in the time window $\tau\in[\tau_2,\tau_3]$ only the ground-state $|0\rangle$ of the free (i.e.  no Liouville potential) quantum mechanics is engaged.  The result is an \emph{exact} reduction of  Eq.~(\ref{eq:quantum-mechanics-four-point-2}) to the product of  two two-point Green functions (\ref{eq:quantum-mechanics}). That is, if only  soft-modes are taken into account,  irreducible four-point correlations are absent,
$ \langle G_{\tau_1,\tau_2} G_{\tau_3,\tau_4} \rangle = \langle G_{\tau_1,\tau_2}\rangle \langle G_{\tau_3,\tau_4} \rangle$.   

Correlations are generated by the massive modes of the theory. The latter were considered in great detail in Refs.~\cite{Polchinski:2016,Maldacena:2016}. These works showed the inclusion of massive modes in the fluctuations of the fields $\Sigma$ and $G$ governed by the action (\ref{eq:action})  is equivalent to the summation of the ladder diagrams. As a result, one obtains the  $1/N$ correction to the four-point Green function
 %\texttt{Avoided the use of the word `irreducible'. The ladder diagrams aren't irreducible in the sense of pert. theory.}
\begin{equation}
\label{eq:one-over-N}
\langle G_{\tau_1,\tau_2} G_{\tau_3,\tau_4} \rangle = \langle G_{\tau_1,\tau_2}\rangle \langle G_{\tau_3,\tau_4} \rangle  \left[1+{1\over N}\, {\cal F}(\chi)\right], 
\end{equation}
where $\chi=\tau_{12}\tau_{34}/\tau_{13}\tau_{24}$ is the conformal ratio. 
The function ${\cal F}(\chi)$ stems from the inversion of the Gaussian part of the action~(\ref{eq:action}), which is equivalent to a summation of ladder diagrams. The presence of soft modes renders the function $F$ nominally divergent, however\cite{Maldacena:2016} that divergence may be separated from the regular massive mode contribution, which under the assumed hierarchy of time scales $\tau_{12},\tau_{34} \ll \tau_{13}, \tau_{24}$ or,  $\chi\ll 1$, may be expanded as ${\cal F}_\mathrm{reg} (\chi)\propto \chi^4+O(\chi^5)$  (see Eq.~(3.90) of Ref.~\cite{Maldacena:2016}). Instead of aiming to regularize  superficially singular soft mode contributions by the $\partial_\tau$ term in the action\cite{Maldacena:2016}, we here follow our master strategy, apply the reparametrization transformation $\tau\to f(\tau)$ to the regular result of the massive mode integration and integrate the $f$-degrees of freedom exactly using the Liouville quantum mechanics. 

In the following we show that this procedure leads produces a correlation contribution to the four point function which for $\tau_{24}\simeq \tau_{13}$ scales as  $N^{-1}|\tau_{13}|^{-3/2}$. This means that even for the different clustering of time arguments the universal long-time $|\tau|^{-3/2}$ law ensues, although it now comes with a small $N^{-1}$ prefactor. 

To show this, we use  $\tau_{12},\tau_{34} \ll \tau_{13}\approx \tau_{24}$ to expand the reparameterized conformal ratio as
\begin{equation}
\chi\to \frac{(f(\tau_2)-f(\tau_1))(f(\tau_4)-f(\tau_3))}{(f(\tau_3)-f(\tau_1))(f(\tau_4)-f(\tau_2))} \sim 
J^{-2}\frac{f'(\tau_1)f'(\tau_3)}{(f(\tau_3)-f(\tau_1))^2}.
\end{equation}
The irreducible part of the four-point function is proportional to $\langle \big(\chi[f]\big)^4\rangle$, where the angular brackets stay for soft-mode averaging. This averaging is essentially equivalent to Eq.~(\ref{eq:2p}) with $p=16$. As a result, it reduces to ${\cal F}_\mathrm{reg} (\chi)$ at short times, while at long times we obtain $\tau_{13}^{-3/2}$.  

\section{Conclusions} 

To summarize, we have shown that the disorder averaged actions describing the SYK model show structural similarity to an `infinite dimensional' nonlinear sigma model: the infinite dimensional symmetry group featuring in the model gets spontaneously broken at mean field level and explicitly broken by time derivatives. The unbroken subgroup, $\mathrm{SL}(2,R)$, is finite-dimensional which leads to the appearance of an infinite dimensional manifold of Goldstone modes. Fluctuations of the latter are damped  solely due to  temporal variations, in a manner described by a nonlinear action which under certain assumptions affords reduction to the `Schwarzian derivative' form~\eqref{eq:soft-action}. Our main observation is that correlation functions computed in this theory can be understood as expectation values of Liouville quantum mechanics, where the bridge between the two theories is established by the nonlinear variable transform~\eqref{eq:phi}. This connection has far-reaching consequences. We apply it to demonstrate that Goldstone mode integrations --- which are nominally divergent in the limit of slow fluctuations at any level of perturbation theory --- lead to finite results when extended over the full manifold. Their integration leads to ultra-universal long time behavior $\tau^{-3/2}$ of arbitrary operator correlation functions. The consequences of these findings within the context of the holographic interpretation of the model remain to be fathomed.

\section{Acknowledgments} 

We thank L. I. Glazman  for bringing the problem to our attention.  A.K. was supported by DOE contract DEFG02- 08ER46482, and acknowledges support by QM2--Quantum Matter and Materials of Cologne University. Work supported by CRC 183 of the Deutsche Forschungsgemeinschaft.

\appendix

\section{}
\label{app:Action}
Here we provide details of derivation of the action~(\ref{eq:expansion}). We start in the time domain where the 2nd-order term in the gradient expansion,
\begin{equation}
S_2[f] =  {N\over 4}\, \mathrm{Tr} (\partial_{\tau} G\partial_{\tau} G) = {N\over 4}\, \int d\tau_1d\tau_2 \, \partial_{\tau_1}\left( G[f]_{\tau_1,\tau_2} \right) \partial_{\tau_2}\left( G[f]_{\tau_2,\tau_1} \right),
\end{equation}
is formulated in terms of the  reparameterized Green functions
\begin{equation}
G[f]_{\tau_1,\tau_2} = f'(\tau_1)^{1/4} G^0\Bigl[f(\tau_1) -f(\tau_2) \Bigr]f'(\tau_2)^{1/4},
\end{equation} 
containing the replica diagonal saddle point solution 
%\begin{equation}
$G^0(\tau) = - {b}\,{\rm sgn}(\tau)|J \tau|^{-1/2}$.
%\end{equation}
We then change the time-integration variables to $t_i = f(\tau_i)$ to arrive at the equivalent representation
\begin{equation}
S_2[f] = {N\over 4}\, \int d t_1d t_2\, \partial_{t_1} \left(   a_{t_1} G^0(t_1-t_2) \,  a_{t_2} \right) \partial_{t_2}  \left(   a_{t_2} G^0(t_2-t_1) \, a_{t_1} \right),  %\quad a_z \equiv \left[ (f^{-1})'(z)\right]^{-1/4}.  
\end{equation}
where we have defined
%\begin{equation}
$a_t \equiv [ (f^{-1})'(t)]^{-1/4}$, 
%\end{equation}
and used the relation $(f^{-1})'(t) f'(\tau)  = 1$ which holds as long as $t=f(\tau)$. Introducing the field $b_t= a_t^2$, switching to center of mass coordinates  $s=\frac 12 (t_1+t_2)$, $t=t_1-t_2$, and integrating by parts, the above action can be cast
into the simpler form~\footnote{We define $f_1(t) \overleftrightarrow{\partial_{t}} f_2(t) \equiv \frac 12 [f_1(t) f_2'(t) - f_1'(t) f_2(t)] $ for any two functions $f_1$ and $f_2$.},
\begin{equation}
S_2[f] = \int \!d s\, d t\, b_{s+t/2}\, \Pi(t)\, b_{s-t/2}, \qquad \Pi(t_1-t_2) =  {N\over 4}\, G^0(t) \overleftrightarrow{\partial_{t_1}} \overleftrightarrow{\partial_{t_2}}  G^0(-t). 
\end{equation}
At this stage it is advantageous to switch to the frequency domain:
\begin{equation}
S_2[f] = \sum_{\omega} b({-\omega}) {\Pi}_\Lambda(\omega) b(\omega),
\end{equation}
where the  Fourier transform of the `polarization operator' $\Pi(t)$ reads
\begin{equation}
\label{eq:Pi_omega}
\Pi_\Lambda(\omega) = -\frac{N}{4}\sum_{|\epsilon| \lesssim \Lambda} \epsilon^2 \, G^0_{\epsilon-\omega/2}  G^0_{\epsilon+\omega/2}, \quad \Lambda \simeq J (f^{-1})'(s).
\end{equation}
It will become clear shortly that the energy integral in the above `polarization
loop' is UV divergent, hence we need to introduce the high-energy cut-off $\Lambda$.
The crucial observation here is that one should preserve the conformal invariance of
the resulting action, meaning that $\Lambda$ should be understood as a function of the `slow' time~$s$.
The self-consistency of this procedure can be easily checked on the example of a
uniform reparametrization $t = \alpha \tau$, where $\alpha$ is a constant. Given that
$\Lambda_0 \simeq J$ is an initial physical cut-off, it should change under rescaling
to $\Lambda = \Lambda_0/\alpha$. Equation~(\ref{eq:Pi_omega}) generalizes this idea
to an arbitrary conformal transformation.

%It is also worth mentioning here that the structure of the polarization operator $\Pi_\Lambda(\omega)$ suggests its interpretation as the inverse propagator of energy fluctuations in the system. 
Evaluating the polarization operator $\Pi_\Lambda(\omega)$ with  logarithmic accuracy we find
\begin{equation}
\label{eq:Pi_log}
\Pi_\Lambda(\omega)  = -\frac{N}{8\pi} \int_{-\Lambda}^\Lambda G^0_{\epsilon-\omega/2}  G^0_{\epsilon+\omega/2} \,\epsilon^2 d\epsilon \simeq \frac{b^2 N}{16 J}  \left( \omega^2 \log(\Lambda/|\omega|) + 4 \Lambda^2\right). 
\end{equation}
Our definitions of $b_s$ and $\Lambda$ imply that the last term in this expression, when transformed back to the time domain, yields an $f$-independent constant,
\begin{equation}
S^0_2[f] \propto \int ds\, b_s^2 \Lambda(s) \propto J^2,
\end{equation} 
which may be omitted. Keeping  only the $\omega$-dependent part of the polarization operator, we obtain the  result
\begin{equation}
S_2[f] =  - \frac{b^2 N}{16 J} \int\!\!\!\!\int dt_1 dt_2 \frac{a_{t_1}^2 a_{t_2}^2}{\left( |t_1 - t_2| + \delta \right)^3}, \qquad \delta \sim \Lambda^{-1}. % = J^{-1}/ (f^{-1})'(s).
\end{equation} 
Restoring the physical time variables, $\tau_i=f^{-1}(t_i)$, and taking into account that $a_t^2 = f'(\tau)^{1/2}$, we finally arrive at the properly regularized soft-mode action~(\ref{eq:expansion})
\begin{equation}
            \label{eq:S2_reg}
S_2[f]=- \frac{b^2N}{16J} \int\!\!\!\!\int d\tau_1 d\tau_2 \,
\frac{f'(\tau_1)^{3/2} f'(\tau_2)^{3/2}}{\left( |f(\tau_1)-f(\tau_2)|  + \delta \right)^{3}}, \quad \delta \sim J^{-1}  f'[(\tau_1 + \tau_2)/2].
\end{equation}
We notice that $S_2[f]$ does not containd  UV (or short-time) divergencies.  In the limit $\tau_1 \to \tau_2$ the denominator of the singular kernel approaches $f'(\tau_1)^3 (|\tau_1-\tau_2| + \delta)^3$ and the $f$-dependence entirely drops out.
This `asymptotic freedom' also guarantees the $\mathrm{SL}(2,R)$ invariance of the action. 
In a similar spirit the Schwarzian action~(\ref{eq:soft-action}) describing  the long-time limit is recovered. Cutting the logarithm  $\log(\Lambda/|\omega|)$ in Eq.~(\ref{eq:Pi_log}) at the scale $\omega \simeq \Delta \, (f^{-1})'(s)$ we obtain the fully local action
\begin{equation}
S_2^{\rm loc}[f] = 2 M \int ds \, \left(\partial_s  b_s \right)^2  \overset{ s = f(\tau)}{=}  - M \int d\tau\, \{f,\tau\}. %\qquad  M=\frac{b^2}{32 J}\,N\log\left({J\over\Delta}\right).
\end{equation} 
with the prefactor $M$ as given in Sec.~\ref{sec:Soft_Mode_Action}.

\section{}
\label{app:Measure}

To deduce the integration measure on the soft-mode manifold we use the fact that
reparamet-rization transformations  $f(\tau)$ form a group and  that the required
measure should be  invariant under group multiplication. Let us denote elements of
this group by $f,g,h,\ldots$, where each is a monotonous function describing a
reparametrization of time, e.g. $\tau\to f(\tau)$. The group multiplication is
defined as subsequent reparameterization, i.e.  $g=f\circ h$ acts as $\tau \to
g(\tau)=h(f(\tau))$. The inverse elements, e.g. $h^{-1}$, are inverse functions,
which are well defined for monotonous reparametrizations. The proper measure
$\mu[g]{\cal D}g$ should keep the path integral invariant under group operations,
i.e. the partition sum
\begin{equation}
Z= \int\! \mu[g] {\cal D}g\, e^{-S[g]} =\int\! \mu[g]{\cal D}g\, e^{-S[g\circ h^{-1}]} 
\end{equation}
is $h$--independent for an arbitrary fixed group element $h$. We now change integration variables as $g=f\circ h$ to find
\begin{equation} 
Z=\int\! \mu[f\circ h]\, \left|\left|\frac{\delta g}{\delta f}\right|\right| \, {\cal D}f\, e^{-S[f]}
\end{equation}
and therefore we conclude that $\mu[f]=  \mu[f\circ h] \,||\frac{\delta g}{\delta f}||_{g=f\circ h} $ at any given $h$.
We may choose $h=f^{-1}$ to find that $\mu[f]=  \mu[1] \,||\frac{\delta g}{\delta f}||_{g=f\circ h; \,h=f^{-1}} $, where the constant $\mu[1]$ may be set to unity. Since $g(\tau)=h(f(\tau))$, we find 
\begin{equation}
\left.\frac{\delta g(\tau)}{\delta f(\tau')}\right|_{g=f\circ h; \, h=f^{-1}}=h'(f(\tau))\Bigl |_{h=f^{-1}} \delta(\tau-\tau') = 
\left(f^{-1} \right)'(f(\tau))  \delta(\tau-\tau')= \frac{ \delta(\tau-\tau')}{f'(\tau)},  
\end{equation}
where in the last equality we have used that for the inverse function $f^{-1}(f(\tau))=\tau$ which in turn implies  
\begin{equation}
\left(f^{-1} \right)'(f(\tau))\, f'(\tau)=1.
\end{equation} 
Thus the invariant measure for the ${\cal D}f$ integration is given by the simple relation
\begin{equation}
\label{eq:f-measure}
\mu[f] =\left|\left|\frac{\delta g(\tau)}{\delta f(\tau)}\right|\right|_{g=f\circ h; \, h=f^{-1}}=\prod\limits_{\tau}\frac{1}{f'(\tau)} .
\end{equation}
Let us further perform the change of variables from $f$ to $\phi$, where $f(\tau)=\int^\tau e^{\phi(\tau')} d\tau'$. 
The corresponding  transformation matrix is of triangular form, $\frac{\delta f(\tau)}{\delta \phi(\tau')}=\theta(\tau-\tau') e^{\phi(\tau')}$, and its determinant therefore given by the product of  diagonal elements,
 $|| \frac{\delta f}{\delta \phi}|| = \prod_{\tau} e^{\phi(\tau)}=\prod_{\tau}f'(\tau)$. Combination with Eq.~(\ref{eq:f-measure}) shows that the  the measure of the ${\cal D}\phi$ integration is {\em flat.} 


%%\bibliographystyle{elsarticle-harv} 
%%\bibliographystyle{elsarticle-num} 

%\vspace{1cm}
%\noindent{\bf References}\\
\bibliographystyle{model1-num-names} 
\bibliography{Bibliography}


%\begin{thebibliography}{00}



%\end{thebibliography}
\end{document}